# Observation of a topologically non-trivial surface state in half-Heusler PtLuSb (001) thin films

J. A. Logan[1], S. J. Patel[1], S. D. Harrington[1], C. M. Polley[2], B. D. Schultz[3], T. Balasubramanian[2], A. Janotti[4], A. Mikkelsen[5], and C. J. Palmstrøm[1,3*]
[1]*Materials Department, University of California-Santa Barbara, Santa Barbara, California 93106, USA*
[2]*MAX IV Laboratory, Lund University, Lund 221 00, Sweden*
[3]*Department of Electrical and Computer Engineering, University of California-Santa Barbara, Santa Barbara, California 93106, USA*
[4]*Department of Materials Science & Engineering, University of Delaware, Newark, Delaware 19716, USA*
[5]*Department of Physics, Lund University, Lund 221 00, Sweden*

**The discovery of topological insulators (TIs), materials with bulk band gaps and protected cross-gap surface states, in compounds such as $Bi_2Se_3$[1] has generated much interest in identifying topological surface states (TSSs) in other classes of materials[2,3]. In particular, recent theory calculations[4–6] suggest that TSSs may be found in half-Heusler ternary compounds. If experimentally realizable, this would provide a materials platform for entirely new heterostructure spintronic devices that make use of the structurally-identical but electronically-varied nature of Heusler compounds. Here, we show the presence of a TSS in epitaxially grown thin films of the half-Heusler compound PtLuSb. Spin and angle-resolved photoemission spectroscopy (ARPES), complemented by theoretical calculations, reveals a surface state with linear dispersion and a helical tangential spin texture consistent with previous predictions[7]. This experimental verification of TI behavior is a significant step forward in establishing half-Heusler compounds as a viable material system for future spintronics devices.**

Half-Heusler compounds are ternary intermetallics (XYZ) which share strong structural similarities to III-V zinc blende binary semiconductors. Crystallographically, these compounds can be thought of as a zinc blende lattice of the X and Z atoms, with additional Y atoms introduced in the octahedral sites (Fig. 1a). Depending on the particular elemental species involved, half-Heusler compounds have a much wider range of electronic structures than III-Vs, due to the important role that the total valence electron count per formula unit has on the Fermi level position[8]. For instance, Heusler compounds have previously been shown to exhibit electronic behaviors such as half-metallic ferromagnetism[9], superconductivity[10], and semiconductivity[11]. Consequently, this flexibility allows for the design of heterostructures of the same crystal structure, but highly varying electronic and magnetic properties. Additionally, this enables the possibility of discovering materials with combined properties, such as topological superconductors which are of interest for the appearance of Majorana fermions[3,12].

Recent first principles calculations have suggested that numerous half-Heusler compounds may exhibit topologically non-trivial behavior[4–7,13]. Focusing on the high average atomic number 18 valence electron per formula unit TI candidates, these materials are predicted to exhibit a zero-gap semiconducting or semimetallic bulk band structure, where, as a result of the interaction between the chemical bonding, crystal field splitting, and spin-orbit coupling, an inversion between the $\Gamma_8$ (p-character) and $\Gamma_6$ (s-character) bands occurs across the Fermi level[4–6]. This band inversion (Fig. 1c) is further predicted to induce the formation of a TSS with linear dispersion, spin-momentum locking, and a Dirac-like crossing analogous to those seen in HgTe/CdTe quantum wells[14] and $Bi_2Se_3$[1]. However, until now, these theoretical predictions have not been experimentally verified in half-Heusler compounds.

For this study, we examine the electronic structure of the half-Heusler compound PtLuSb (001). Samples were grown by molecular beam epitaxy (MBE) and characterized by reflection high-energy electron diffraction and x-ray diffraction to ensure high sample quality (see supplemental information). First-principles calculations[5,6] predict PtLuSb to lie at the border between normal and inverted band ordering with a zero-gap semiconducting band structure. Previous experimental studies on bulk single crystals[15,16] and thin films[17] have confirmed that PtLuSb has the expected zero-gap density of states, but have not measured the momentum-resolved electronic band structure. Consequently, we utilize spin-integrated and spin-resolved ARPES to directly probe the electronic band structure and search for the predicted TSS. Here, we point-by-point identify all of the relevant signatures of a TSS: first, no out-of-plane dispersion should be observed due to its confined surface nature; second, a linear in-plane dispersion should be apparent with no evidence of partner bands; and third, spin-momentum locking should be evident[2]. Experimental ARPES data are directly compared with new *ab-initio* calculations that further supports the results.

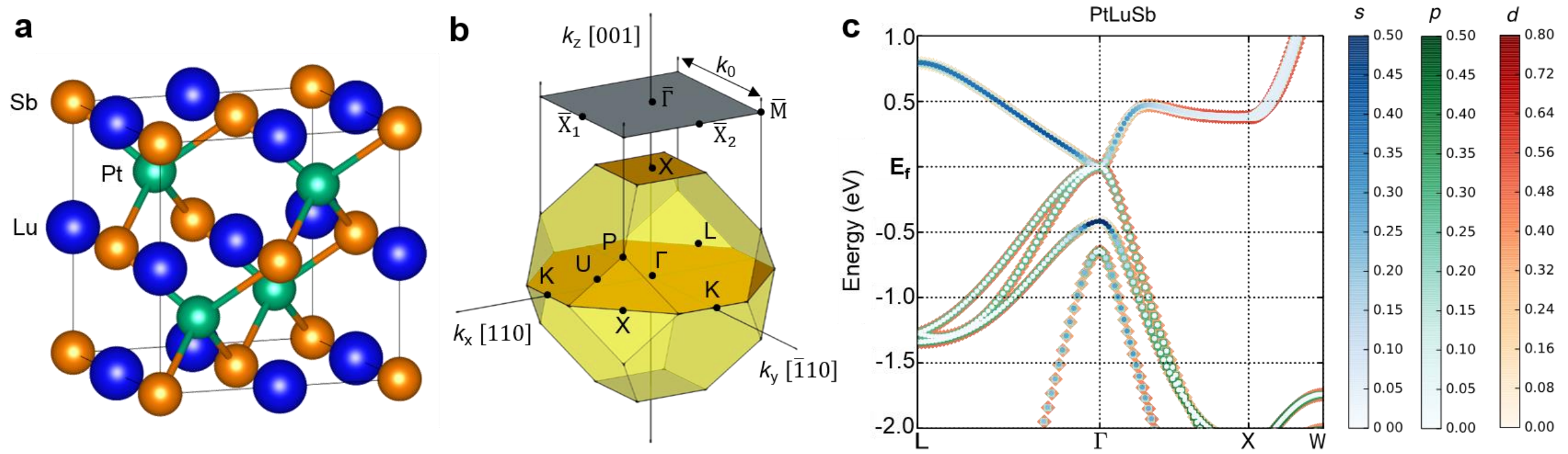

*Figure 1.* ***a****, Half-Heusler $C1_b$ crystal structure consisting of three interpenetrating face-centered-cubic sub-lattices. For PtLuSb the platinum, lutetium, and antimony atoms are denoted by the green, blue, and orange circles, respectively.* ***b****, Half-Heusler bulk BZ with the (001) surface BZ projection. The high-symmetry surface BZ points are defined as $\bar{\Gamma}$ (0, 0), $\bar{X}_1$ ($k_0$/2, 0), $\bar{X}_2$ (0, $k_0$/2), and $\bar{M}$ ($k_0$/2, $k_0$/2) where, for PtLuSb (001), the surface unit momentum $k_0 = 2\pi(\sqrt{2}/a_0) = 1.376$ Å$^{-1}$.* ***c****, First-principles calculated bulk electronic band structure of PtLuSb with corresponding band character shown. A band inversion can be clearly seen at the bulk Γ point, where the s-like band is below the p/d-like band.*

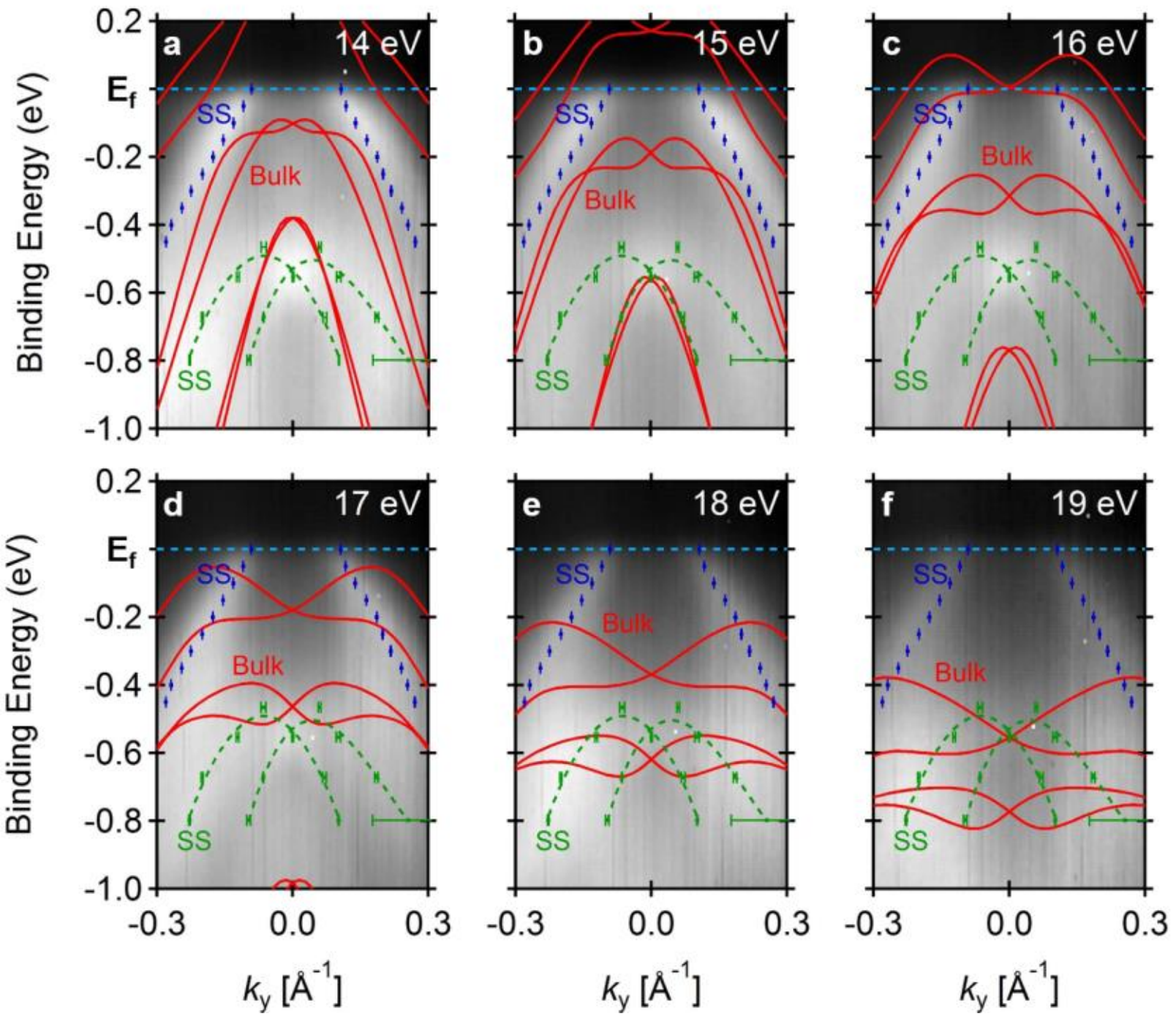

*Figure 2. In-plane ARPES dispersion maps, and calculated bulk band structure (red) with overlaid extracted surface state positions (blue and green) along the $\bar{\Gamma} - \bar{X}_2$ directions for incident photon energies of (****a-f****) 14-19 eV, respectively. The maps highlight the movement of the bulk bands as a function of changing $k_z$ as well as the presence of three surfaces states. The theory calculation Fermi level has been shifted -0.350 eV to align with the experimental Fermi level position.*

To begin, we examine the in-plane ARPES dispersion maps for various photon energies (Fig. 2) in order to observe the bulk band motion relative to any surface states. Changing the incident photon energy results in an adjustment in $k_z$ enabling bulk bands, which disperse with $k_z$, to be easily distinguished from surface states, which do not disperse in $k_z$, at normal emission ($k_x$, $k_y$ = 0). For PtLuSb (001), an incident photon energy of 13 eV corresponds to the bulk Γ point at the Fermi level while an incident photon energy of 26 eV corresponds to the bulk X point at the Fermi level, based on the bulk band periodicity. By comparing spectra taken at 14-19 eV, we observe two sets of high intensity bands which are non-dispersive with photon energy (position extractions from Gaussian peak fits are overlaid on the calculated bulk bands): one set which crosses the Fermi level (blue) and one set approximately 0.5 eV below the Fermi level at $\bar{\Gamma}$ (green). Additionally, we note the presence of several weaker intensity bands which appear to move downward in binding energy as $k_z$ moves toward the bulk X point (higher photon energies), consistent with the projected bulk bands (red). Examining these non-dispersive states in more detail, we first note the lower state's similarity to a Rashba split hole band. This split hole band is comparable, both in terms of shape and binding energy, to features seen in ARPES measurements of cleaved bulk PtLuBi (111) and PtGdBi (111) half-Heusler crystals[18]. Consequently, it is reasonable to assume that this lower surface state arises due to a common crystallographic or chemical feature shared by a variety of high-Z half-Heusler compounds. Second, we note that the upper non-dispersive state is qualitatively similar to the lower half of a Dirac cone, matching with the expectation for a TSS.

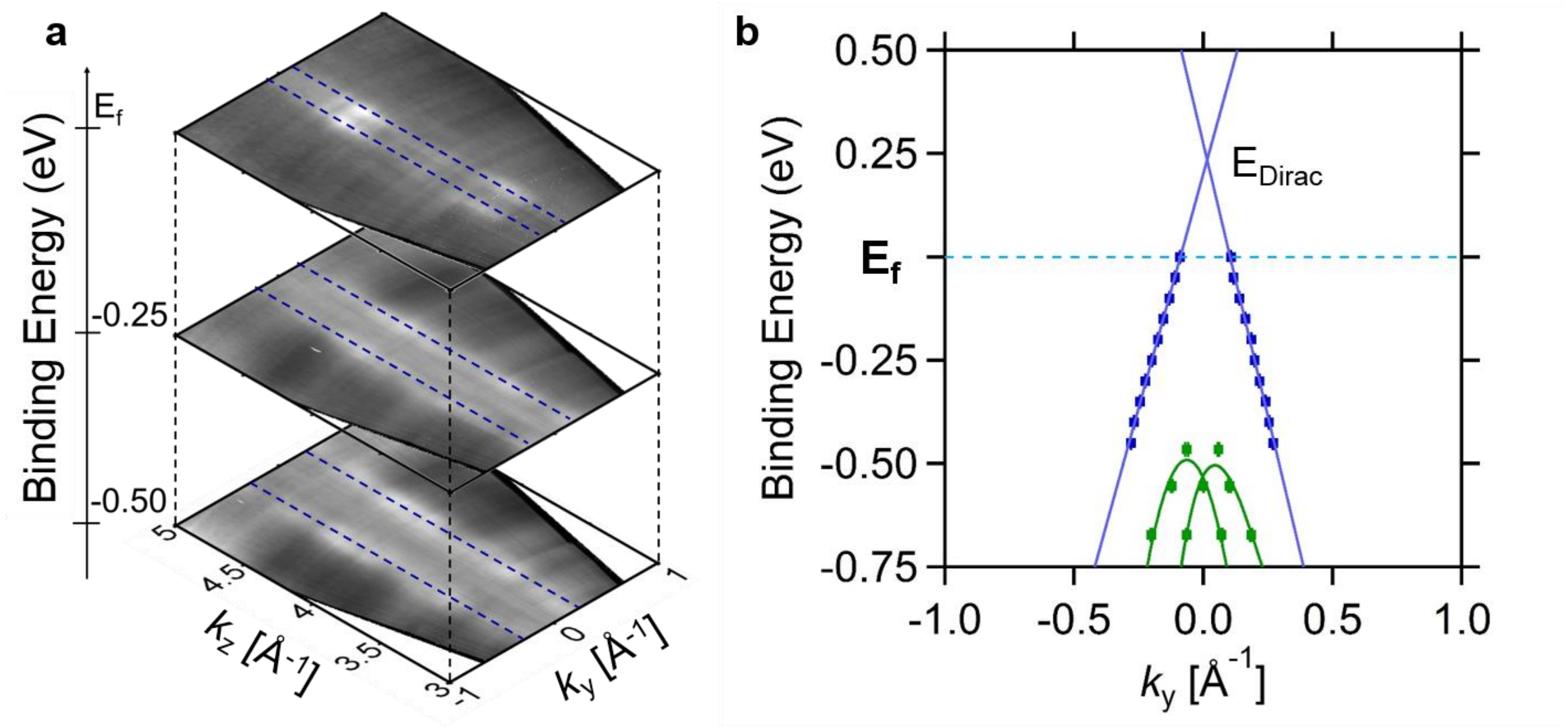

*Figure 3. **a**, Intensity as a function of out-of-plane and in-plane momenta for various binding energy slices. The upper surface state position is marked in blue. An absence of $k_z$ dispersion can be observed, consistent with the behavior of a surface state. Intensity modulations along $k_z$ are the result of surface state resonance. **b**, By tracking the surface state's position through binding energy, the predicted linear in-plane dispersion can be seen and a Dirac point can be extrapolated to 0.235 ± 0.010 eV above the Fermi level. Error bars corresponding to one standard deviation for a Gaussian peak fit fall within the data points.*

To prove that this observed Dirac-like state arises from the sample's surface, we scan the incident photon energy over a wide range at normal emission (corresponding to sweeping along bulk Γ-X for several BZs) and simultaneously examine the binding energy dependence along the $\bar{\Gamma}-\bar{X}_2$ direction (Fig. 3a). We immediately note a lack of $k_z$ dispersion, in agreement with our initial in-plane observations (Fig. 2) and consistent with the behavior of a surface state. Intensity modulations appear within the surface state due to resonance near bulk Γ points. Furthermore, by extracting the average surface state position for various binding energies, a linear dispersion can be seen (Fig. 3b); extrapolating this Dirac-like dispersion upward yields a crossing 0.235 ± 0.010 eV above the Fermi level. In agreement with a previous experimental study[17] where Hall measurements of similar films revealed ~1 x $10^{20}$ $cm^{-3}$ p-type carriers, we observe that the Fermi level falls within the valence band. This is in contrast with theory calculations which predict the Fermi level to fall at the valence band maximum, suggesting that there may be a low-energy defect which induces p-type doping, similar to antimony antisite disorder in III-Sb semiconductors[19].

To check for other surface states and obtain a more complete picture of the Fermi surface (FS), a higher photon energy can be used to increase the accessible reciprocal space area. Seen in Fig. 4a, it is clear that the FS is more complicated than existing calculations suggest. The sample's (1x3) surface reconstruction, seen in low-energy electron diffraction (Fig. 4b), contributes an additional replica state with a three-fold periodicity in the $\bar{\Gamma}-\bar{X}_1$ direction. The intensity difference between the first and second $\bar{\Gamma}$ points is consistent with considerations of the bulk Brillouin zone projections (see supplemental information). At this photon energy, the first $\bar{\Gamma}$ point projects from near the bulk X point, where the bulk bands are far from the Fermi level, while the second $\bar{\Gamma}$ point projects from near the bulk Γ point of an adjacent BZ, leading to overlap of surface and bulk bands and greatly increased intensity. Focusing on the potential TSS near $\bar{\Gamma}$ more closely, we measure constant energy contours at photon energies of 16 eV (Fig. 4d-g) and 18 eV (Fig. 4h-k). Much like in $Bi_2Te_3$ (111)[20,21] and W (110)[22] where warping is seen, the constant energy contours are anisotropic, eventually resembling the square of the (001) surface BZ at binding energies far from the Dirac point. Furthermore, the linear state's rectangular FS is convoluted with features from the three-fold state leading to the appearance of distortions, particularly along the shorter edge at lower binding energies. By examining the second-derivative maps we note the presence of several weak features closer to $\bar{\Gamma}$ than the surface state with linear dispersion, which disappear as photon energy is increased to 18 eV. Consequently, we attribute these states to the bulk bands, which agrees well with our calculated bulk FS structure.

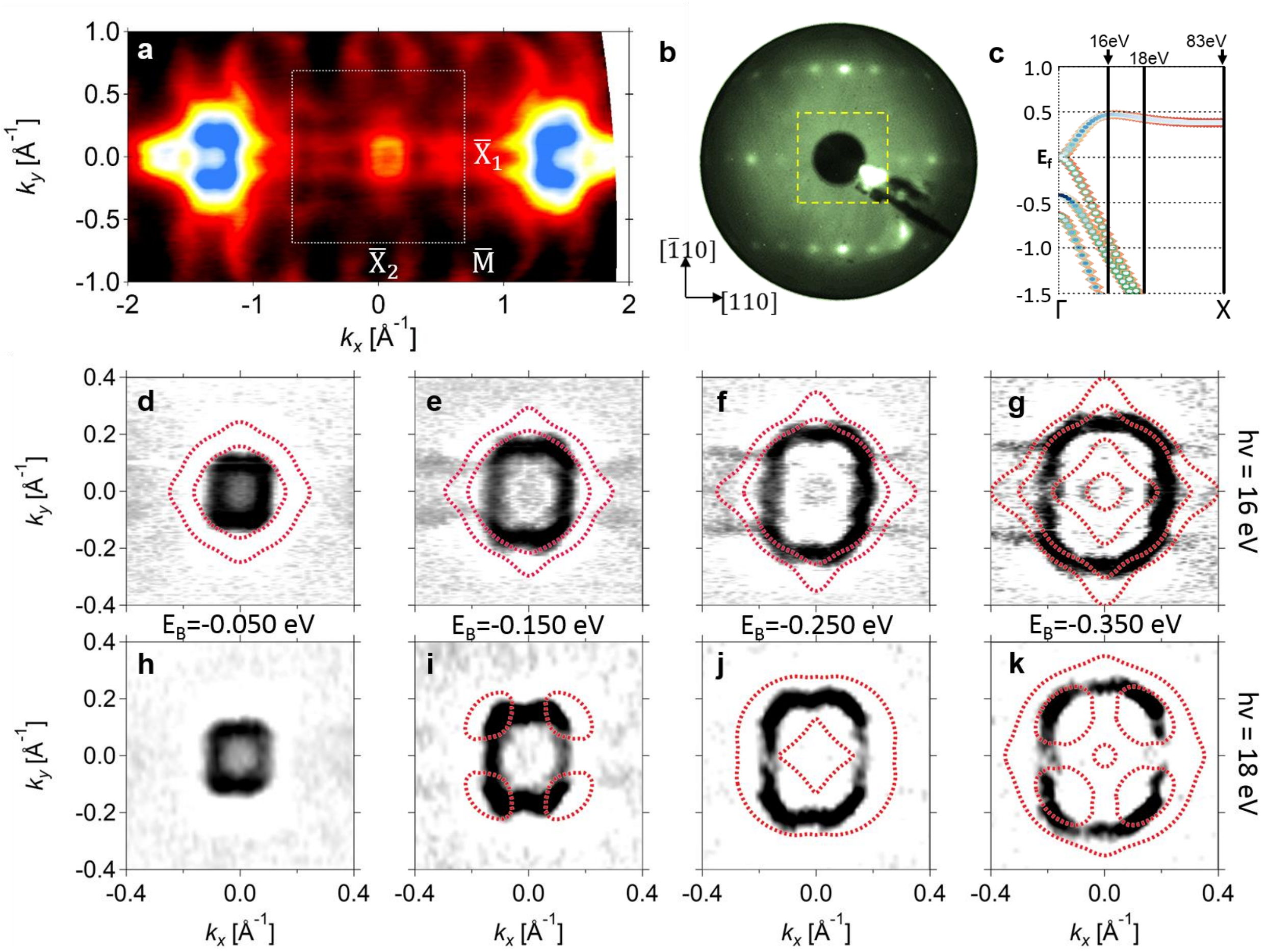

*Figure 4. **a**, The observed constant binding energy surface of PtLuSb (001) with an incident photon energy of 83 eV (the bulk X point) and binding energy of -0.075 eV. The higher photon energy increases the amount of accessible reciprocal space at a cost of resolution. A three-fold periodic surface state can be seen in addition to the topological surface state. The second-order $\bar{\Gamma}$ points ($k_x = \pm k_0$, $k_y = 0$) have greatly increased intensity because they project from near the bulk Γ point of the adjacent BZ enabling both surface and bulk states to be seen. **b**, 75 eV LEED pattern of the sample's surface revealing a (1x3) surface reconstruction. **c**, Schematic diagram highlighting $k_z$ position of the 16 eV, 18 eV, and 83 eV FS maps. Finally, high-resolution second-derivative FS maps near $\bar{\Gamma}$ for an incident photon energy of (**d-g**) 16 eV and (**h-k**) 18 eV for various binding energies showing a complex Fermi-surface shape, overlaid with the calculated bulk structure. The theory calculation Fermi level has been shifted -0.350 eV to align with the experimental Fermi level position.*

Lastly, we conduct spin-ARPES to map the spin texture of the surface states closely surrounding $\bar{\Gamma}$. By examining the spin-polarization and energy distribution curves (EDCs) for various points on the FS (Fig. 5e-g), it becomes clear that the surface state with linear dispersion has a strong spin texture with either side having opposite polarization. Furthermore, we observe a counter-clockwise helical tangential spin-texture with minimal radial components, similar to that seen in the lower cone of $Bi_2Se_3$[1] and consistent with theoretical calculations for a TSS in half-Heusler compounds with positive spin-orbit coupling[7]. We also observe that the surface state at lower binding energy is spin-split, consistent with the expectation for a Rashba-like split hole band. In both cases, while our calculations suggest the bulk bands may also exhibit spin-splitting, the low detection efficiency for spin-ARPES causes only the highest intensity states to be measured. Consequently, when we consider the intensity difference seen between the bulk and surface states in Fig. 2, the observed spin-signal must be dominated by surface state behavior.

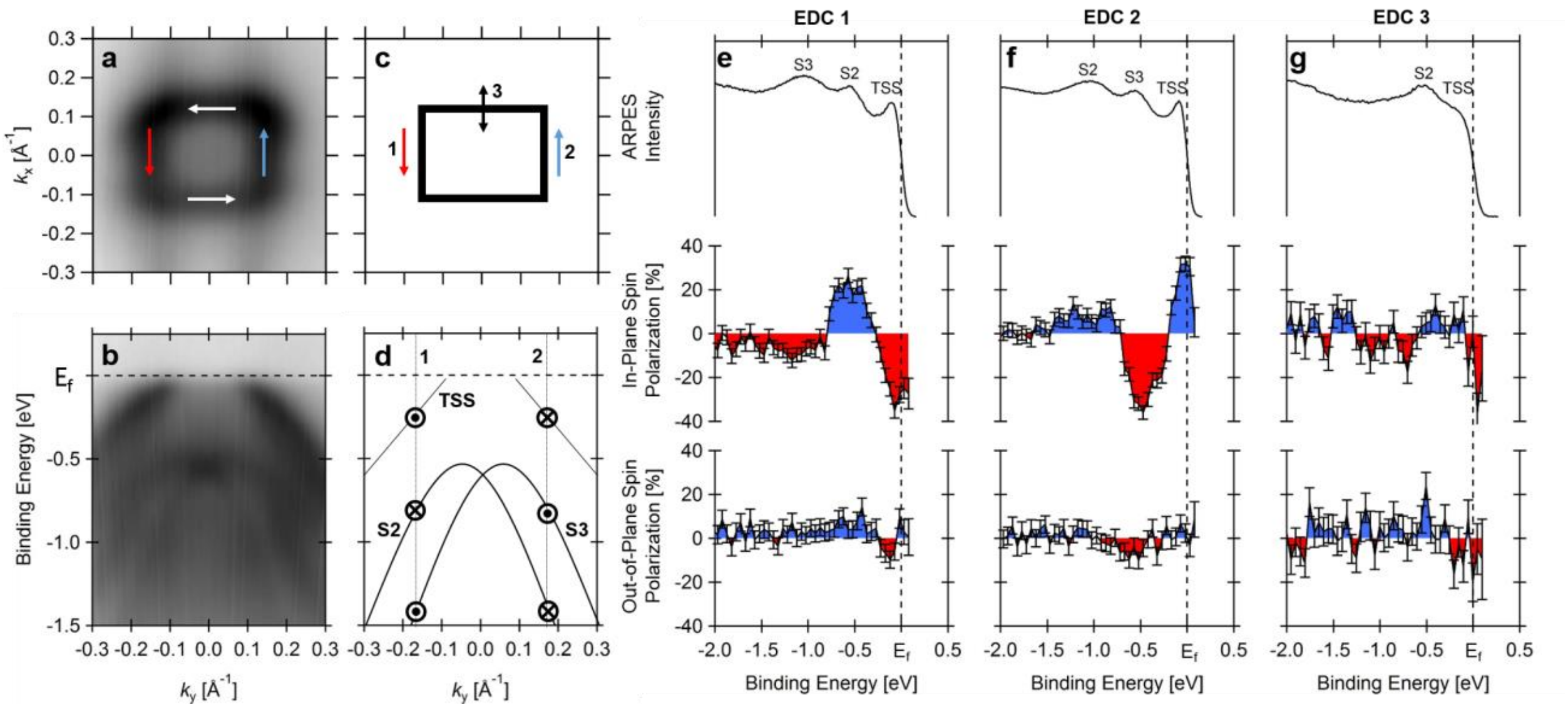

*Figure 5.* ***a****, Constant binding energy surface with a photon energy of 16 eV highlighting the surface state with linear dispersion.* ***b****, In-plane ARPES dispersion along the $\overline{\Gamma} - \overline{X_2}$ direction emphasizing the position of the surface state with linear dispersion and the lower Rasbha-like surface state.* ***c****, Schematic FS diagram depicting the location of EDC 1-3.* ***d****, Schematic in-plane ARPES dispersion with the measured in-plane spin-polarization marked. The surface state with linear dispersion shows strongly opposing polarizations on either side. Furthermore, the spin-polarization for Rashba-like surface state alternates as expected.* ***e-g****, Energy distribution curves and spin polarizations for the noted locations. For in-plane spin-polarization, positive (blue) spin-polarizations correspond to polarization toward positive $k_x$, negative (red) spin-polarizations correspond to polarization toward negative $k_x$. For out of-plane spin-polarization, positive (blue) spin-polarizations correspond to polarization toward positive $k_z$, negative (red) spin-polarizations correspond to polarization toward negative $k_z$. Consequently, considering EDC 1-3 we note the presence of a counter-clockwise helical spin texture with minimal radial component (see supplemental information for details of the analysis and additional spin and circular dichroism measurements).*

Although the experimental data is well explained by a TSS, a possible alternative explanation is that the upper surface state is instead a trivial surface state with unequal contributions of bulk (Dresselhaus) and structural (Rashba) inversion asymmetries, similar to that shown by Ganichev and Prettl[23]. However, several observations argue against such an interpretation. First, a Rashba/Dresselhaus combination state requires the presence of a partner band with opposite spin polarization. Examining our spin-ARPES data, we do not observe any such behavior. Furthermore, in the event that partner bands were too close together to resolve in momentum space, the observed spin-polarization would be very weak or non-existent, in contradiction with the measured data. Second, examining the spin-integrated ARPES data, we note that while there is some residual intensity between the Fermi level and the lower surface state with an incident photon energy of 16 eV, the intensity drops rapidly as $k_z$ is moved toward the bulk X point, consistent with the bulk band motion. Third, the surface state extraction fits to a linear dispersion with high accuracy. This is in contrast with the expectation for a trivial Rashba/Dresselhaus surface state, where a purely linear dispersion would not be expected[23]. Finally, the observed surface state behavior corresponds well both with theory in literature as well as our own first-principles calculations. Therefore, we conclude that no partner bands are present and that consequently the observed surface state is not a trivial Rashba/Dresselhaus surface state.

In summary, we report the first direct experimental observation of a topologically non-trivial surface state in the half-Heusler material system. Our spin-ARPES data of PtLuSb (001), supported by first-principles calculations, reveals a topological state that has a counter-clockwise helical spin polarization and extrapolated Dirac point 0.235 ± 0.010 eV above the Fermi level. It is important to note that while unstrained PtLuSb is a zero-gap semiconductor, it is the first half-Heusler compound to be shown to exhibit topologically non-trivial surface states. With this experimental verification, it opens the door for the numerous other compounds in the Heusler family that have been suggested to possess the necessary bulk band-inversion, some of which have non-zero bulk bandgaps. This continues to highlight the multifunctional nature of Heusler compounds by demonstrating the presence of another unique electronic structure, potentially enabling a range of new heterostructure devices that combine the widely varied electronic properties of these materials without changing crystal structure. Consequently, Heusler compounds are an exciting and promising direction for further exploration to find new topologically non-trivial materials as well as for the potential to discover materials with new electronic structures such as topological superconductors.

## METHODS

**Growth approach.** Samples consisting of 18 nm of relaxed PtLuSb on 5 monolayers of GdSb on relaxed ~1x10$^{18}$ cm$^{-3}$ beryllium doped $Al_{0.1}In_{0.9}Sb$ on p+ GaAs (001) substrates with a ~200 nm antimony cap were grown by molecular beam epitaxy (MBE) with a system base pressure of 1x10$^{-10}$ Torr. 500 nm thick ~1 x 10$^{18}$ cm$^{-3}$ beryllium doped GaAs buffer layers were first grown under an $As_4$ overpressure to form a smooth (2x4) surface reconstruction. $Al_{0.1}In_{0.9}Sb$ was then nucleated directly on the GaAs at 380°C by first soaking the GaAs surface with $Sb_2$ for 10 seconds and then beginning co-deposition. Growth was continued as the substrate was quickly heated to 450 °C in order to smooth the surface and prevent excess antimony from sticking, similar to the technique developed by Davis *et al*[24]. $Al_{0.1}In_{0.9}Sb$ layers were terminated by an asymmetric (1x3) surface reconstruction after 300 nm of growth. 5 monolayers of GdSb, calibrated by Rutherford backscattering spectrometry (RBS) of elemental films and RHEED oscillations[25], were then grown to provide a diffusion barrier[26] between the PtLuSb and the $Al_{0.1}In_{0.9}Sb$ buffer structure. Samples were then transferred *in-situ* to a seperate MBE system for PtLuSb growth. Lutetium and antimony were evaporated from effusion cells while platinum was evaporated from an electron beam evaporator. Beam fluxes were calibrated by a combination of RBS and RHEED intensity oscillations for lutetium and antimony and by a quartz crystal microbalance for platinum to obtain a 1 lutetium : 1 platinum : 1.2 antimony atomic flux ratio. Nucleation of the PtLuSb was initiated with a low temperature shuttered sequential monolayer growth technique followed by co-deposition at 380°C utilizing the procedure developed by Patel *et al*[17,27]. At this temperature, antimony is self-limited, similar to the $Al_{0.1}In_{0.9}Sb$ growth. RHEED patterns showed either streaky (1x3) or c(2x2) surface reconstructions depending on antimony overpressure and substrate temperature. Finally, to prevent oxidation during *ex-situ* transfer to the MAX-lab synchrotron facility, a ~200 nm antimony capping layer was deposited at 100 °C. X-ray diffraction analysis shows clear thickness fringes consistent with the expected growth rate.

**Experimental approach.** ARPES and spin-ARPES measurements were taken at beamlines i4 and i3, respectively, at the MAX-lab synchrotron facility in Lund, Sweden. Before measurement, samples were heated to ~390 °C whereupon the antimony capping layer was desorbed, as confirmed by low-energy electron diffraction (LEED) and examination of shallow core levels by angle-integrated photoemission spectroscopy. Spin-integrated ARPES was conducted at 100 K to reduce thermal broadening while spin-resolved ARPES was conducted at 300 K with vacuum conditions of 1x10$^{-10}$ Torr. The observed band structures were stable throughout the entire measurement duration.

**Theoretical approach.** The calculations are based on the density functional theory (DFT) within the generalized gradient approximation (GGA) as implemented in the VASP code[28]. The interactions between the valence electrons and the ion cores were described using projected augmented wave potentials. A planewave basis set with cutoff of 300 eV was employed, and a special *k*-point mesh of 12x12x12 was used for the face-centered cubic primitive cell. The calculated $a_0$ = 6.512 Å is less than 1% larger than the experimental value of 6.457 Å. A *k*-point mesh 8x8x6 was used in the calculations for the 6-atom tetragonal cell ($\sqrt{2}/2a_0$ x $\sqrt{2}/2a_0$ x $a_0$), where $a_0$ is the equilibrium lattice parameter. The electronic band structure calculations were performed including spin-orbit interaction.

**ACKNOWLEDGEMENTS**

We thank the beamline staff and scientists at MAX-lab for their support. Additionally, we thank J. Osiecki for his previous work on ARPES analysis code[29]. Funding for the experimental work was provided in part by the Office of Naval Research, the Army Research Office, and the National Science Foundation Materials Research and Science and Engineering Center. Funding for the theory work was provided by the U.S. Department of Energy. Travel support was provided by the IMI Program of the National Science Foundation. We also acknowledge the use of facilities within the LeRoy Eyring Center for Solid State Science at Arizona State University.

# Supplemental Material for "Observation of a topologically non-trivial surface state in half-Heusler PtLuSb (001) thin films"

J. A. Logan[1], S. J. Patel[1], S. D. Harrington[1], C. M. Polley[2], B. D. Schultz[3], T. Balasubramanian[2], A. Janotti[4], A. Mikkelsen[5], and C. J. Palmstrøm[1,3*]
[1]*Materials Department, University of California-Santa Barbara, Santa Barbara, California 93106, USA*
[2]*MAX IV Laboratory, Lund University, Lund 221 00, Sweden*
[3]*Department of Electrical and Computer Engineering, University of California-Santa Barbara, Santa Barbara, California 93106, USA*
[4]*Department of Materials Science & Engineering, University of Delaware, Newark, Delaware 19716, USA*
[5]*Department of Physics, Lund University, Lund 221 00, Sweden*
(Dated November 13, 2015)

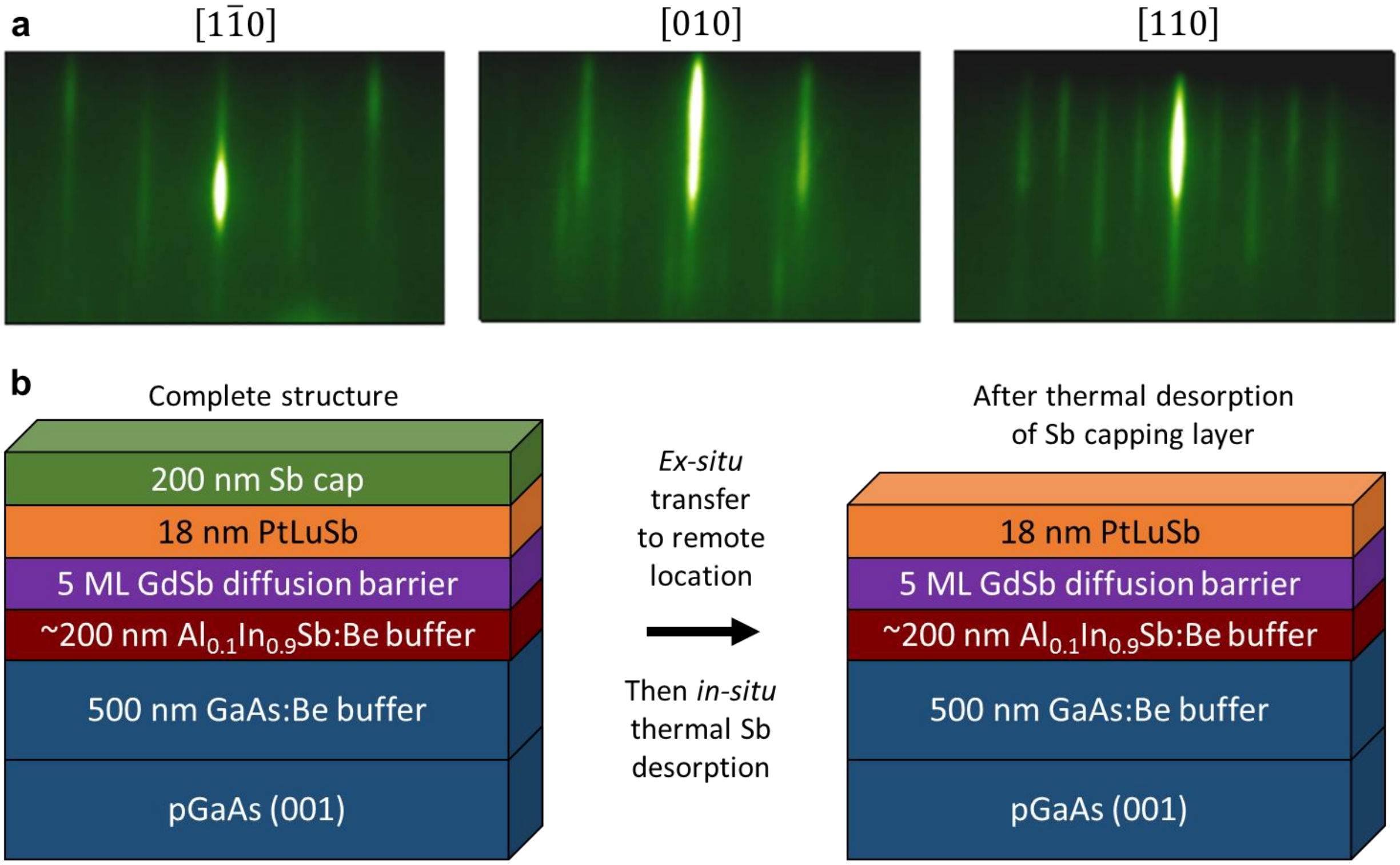

*Figure S1.* ***a****, Reflection high-energy electron diffraction patterns of the 18 nm PtLuSb film prior to deposition of the protective antimony capping layer showing a sharp streaky pattern consistent with a smooth, high quality film.* ***b****, Schematic diagram of the sample structure before and after thermal desorption of the antimony capping layer. This capping layer prevents the PtLuSb from oxidizing during ex-situ transfer between the growth system and the ARPES system.*

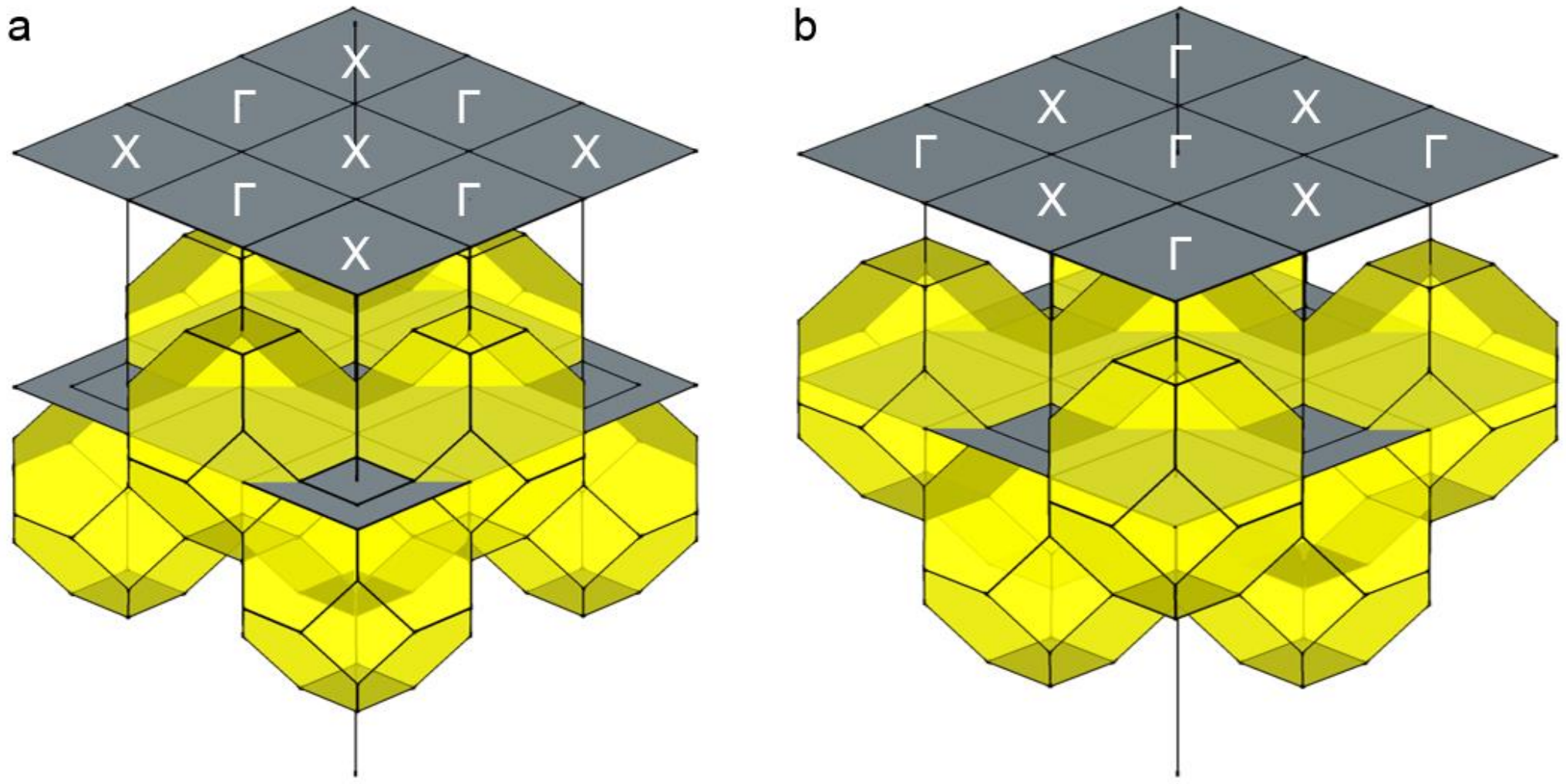

*Figure S2. Bulk to (001) surface Brillouin zone projections including neighboring bulk zones centered at,* ***a****, the bulk X point and* ***b****, the bulk Γ point. Labels correspond to the bulk point projecting to $\bar{\Gamma}$ for the respective surface Brillouin zone.*

## Characterization by Spin-ARPES and Circular Dichroism

Spin ARPES and circular dichroism measurements were performed at the I3 beamline of the MAX-IV laboratory, Sweden[1]. This beamline employs a Scienta R4000 hemispherical electron analyzer, configured to output either to a 2D MCP detector for spin-integrated measurements or to a mini-Mott spin detector. The latter permits simultaneous detection of two spin components by measuring the intensity difference between two channeltron detector pairs, one sensitive to in-plane polarization (along the analyzer slit direction) and one sensitive to out-of-plane polarization.

After scaling the channeltrons by known, constant sensitivity factors and subtracting a dark-count background, spin polarization was computed as[2]:

$$(1)\ Polarization = \frac{I_A - I_B}{S * (I_A + I_B)}$$

where the value of the Sherman function S = 0.15, and $I_A$ and $I_B$ are the counts recorded by the channeltron pair. Error bars are based on counting statistics, and computed as:

$$(2)\ \Delta Polarization = \sqrt{\frac{1}{N * S^2}}$$

where N is the total intensity measured by a given detector pair.

In Figure S3, we show additional spin-resolved spectra acquired at a photon energy of 18eV. Despite changing the photon energy, the spin texture observed in the main manuscript is unchanged, confirming that the observed spin polarizations originate from the surface bands, rather than the bulk bands.

In Figure S4a we show a standard ARPES spectrum acquired with vertically polarized light at a photon energy of 16eV. By combining similar spectra taken with left (LCP) and right circularly polarized (RCP) light, we obtain the corresponding circular dichroism spectrum (Figure 4c). The surface states at the valence band show a strong dichroic signal of up to 40% (the beamline polarization purity at this photon energy is estimated to be approximately 85%). Figure S4b illustrates the alignment of the sample for these spectra, whilst Figure S4d depicts the experimental geometry of the beamline. If interpreted with caution, circular dichroism in topological insulators can reflect the spin-orbit texture and hence provide a higher efficiency alternative to direct spin-resolved measurements[3,4].

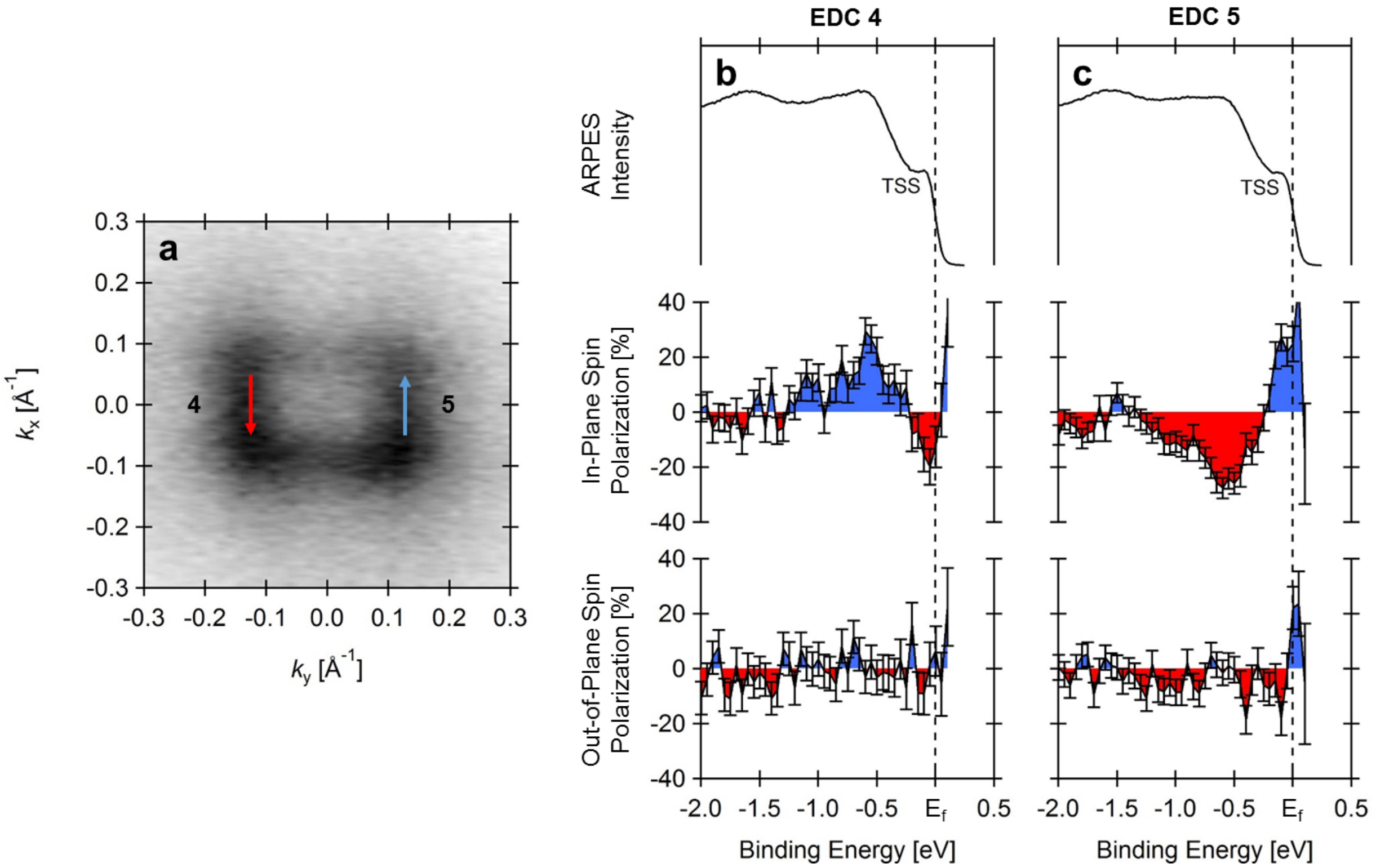

*Figure S3.* ***a****, Constant binding energy surface with a photon energy of 18 eV highlighting the surface state with linear dispersion and noting the measured spin-polarization orientations.* ***b-c****, Energy distribution curves and spin polarizations for the noted locations. For in-plane spin-polarization, positive (blue) spin-polarizations correspond to polarization toward positive $k_x$, negative (red) spin-polarizations correspond to polarization toward negative $k_x$. For out of-plane spin-polarization, positive (blue) spin-polarizations correspond to polarization toward positive $k_z$, negative (red) spin-polarizations correspond to polarization toward negative $k_z$. The observed spin-polarization spectra are unchanged compared to those taken at 16 eV.*

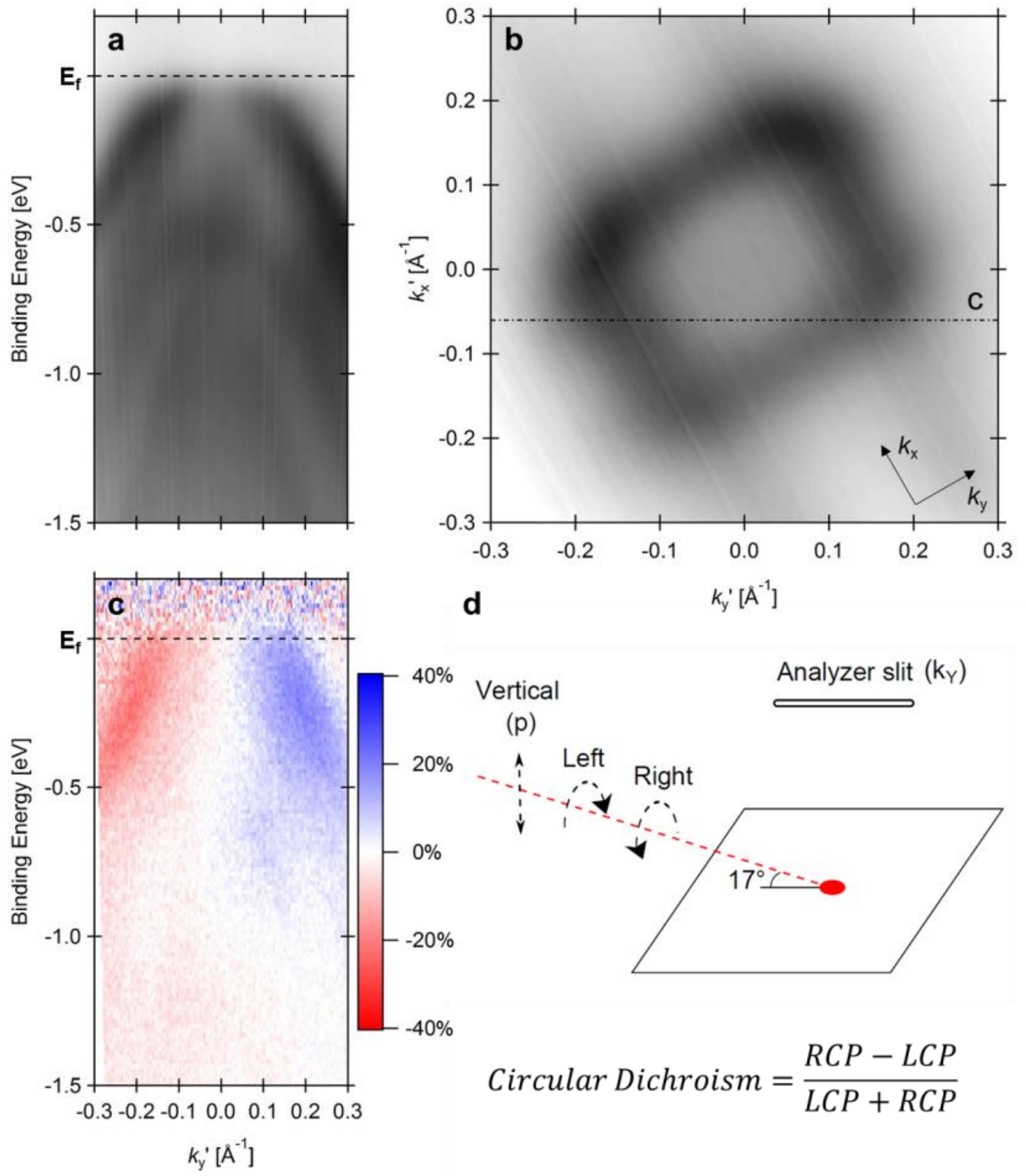

*Figure S4. **a**, 16 eV ARPES spectrum with vertically polarized light. **b**, Corresponding FS, illustrating the sample alignment and noting the position of spectra (a) and (c). **c**, Circular dichroism spectrum showing the TSS has a strong dichroic signal of up to 40% (the polarization purity at the I3 beamline at this photon energy is estimated to be approximately 85%). **d**, Experimental geometry of the beamline illustrating the incident beam angle and polarization orientations.*

## Supplementary References